\documentclass{elsart}
\usepackage{graphicx}
\usepackage{amsfonts}
\usepackage{amsmath}

\journal{Eur. Phys. J. B}
\begin{document}

\begin{frontmatter}

\title{Finite Size Effects on the Optical Transitions in  Quantum Rings under a Magnetic Field}

\author[CA]{Tatyana V. Bandos},
\author{Andr\'{e}s Cantarero},
and
\author{Alberto Garc\'{\i}a-Crist\'{o}bal}
\address{Material Science Institute, Universitat de Val\`{e}ncia, PO Box 22085,
46071 Valencia, Spain.}
\thanks[CA]{E-mail address:
Tatyana.Bandos@uv.es.}

\begin{abstract}
We present a theoretical  study of the energy spectrum of single electron and hole states in
quantum dots of annular geometry under a high magnetic field along the ring axis in the frame of
uncorrelated electron-hole theory. We predict the periodic disappearance of the optical emission of
the electron-hole pair as the magnetic field increases, as a consequence of the finite height of
the barriers.  The model has been applied to semiconductor rings of various internal and external
radii, giving as limiting cases the disk and antidot.
\end{abstract}

\end{frontmatter}


\section{Introduction}

The Aharonov-Bohm (A-B) oscillations \cite{AB59} in mesoscopic
metallic and semiconductor rings have attracted much interest and
have been studied both experimentally \cite{levy90,mail93,fuhrer01}
and theoretically
\cite{buttiker,byers,gefen,chakra92,chap95,raikh00}. The A-B quantum
interference effect in such systems arises from the phase shift
accumulated by the wave function of a charged particle moving in a
ring pierced by an external magnetic field.

Recently, magneto-transport experiments  have been complemented by
the detection of the A-B signature through optical properties of
mesoscopic rings \cite{bayer04,govorovII04}. Since it has become
possible to manufacture self-assembled semiconductor quantum dots
with ring shape there have been considerable efforts to develop
spectroscopic tools capable to detect the energy spectrum  in a
magnetic field because of their high optical
quality~\cite{warburton04}. Ring-like properties of few electrons in
self-assembled quantum dots under a magnetic field have been studied
by far-infrared spectroscopy~\cite{lorke00}.

The main physics of  the A-B optical oscillations  for magneto-exciton was predicted in quantum
rings (QRs) \cite{chap95,gov97}. The elementary neutral excitation, the so-called {\it neutral}
exciton, is  a bound state of a hole  in a filled valence band and an electron in an otherwise
empty conduction band of the host semiconductor. The A-B effect in a semiconductor quantum ring for
charged particle has recently been studied in optical experiments~\cite{bayer04}, while for neutral
exciton it has not been observed yet in type-I QRs.

Govorov et al.~\cite{gov02,ull02,ph02} have considered the possibility to detect optically an A-B
signature for a polarized electron-hole ({\it e-h})  pair in a quantum ring of  finite width. An
essential feature of their study is the radial polarization, incorporated by the modeling of the
neutral exciton  with and without Coulomb interaction in the  QR as two  concentric parabolic rings
for the electron and hole. In both  limiting cases of strong and weak interaction, the radial
polarization allows for the optical manifestation of the topological A-B effect for neutral exciton
in  a finite width ring, in contrast with the exponentially small amplitude of the A-B oscillations
of the ground state energies in "zero" width quantum rings~\cite{chap95,raikh00,ulloa01,zhu01}.
Recently, the A-B effect was studied for spinless interacting hole and electron within the frame of
attractive Hubbard model on the discrete lattice of the few rings in vacuum \cite{palm05}. One can
state that this  paper, like \cite{chap95,raikh00,ulloa01,zhu01},  takes into account {\it e-h}
Coulomb interaction, but by modeling it by the exactly soluble Hubbard Hamiltonian ({\it e.g.} in
\cite{raikh00} the delta-function potential is used instead). The conclusions in \cite{palm05}
include that the A-B effect is suppressed when the QR circumference becomes large, whereas it is
due to the large width in accordance to \cite{zhu01}. It was also noted in \cite{palm05} that the
phenomenon in the two-dimensional (2D) model is analogous to one in the one-dimensional (1D)
system. This is the place to notice that both A-B and Aharonov-Casher \cite{AC84} effects in the
N--fermion Hubbard model with attraction at the  1D ring lattice under a radial electric field have
been studied in \cite{zvyag93} (see also \cite{zvyag96}). The enhancement of A-B effect due to
external electric field  in the 2D ring was studied in \cite{maslov03}.

The interplay between the excitonic radial polarization and Coulomb interaction was analyzed in
\cite{ulloa05,bart06}, assuming that the hole and electron are restricted to 1D concentric rings.
Here we consider  non--interacting  electron and hole both in the  2D ring immersed in a matrix of
host semiconductor. Such a simple model allows us to study tunneling effect through finite height
potential (instead of infinite potential as, {\it e.g.} in \cite{ulloa01}). The height of the hard
wall potential influences the angular momentum transitions \cite{jan01} which are one of the
subjects of our main interest. It has been proposed that type-II quantum dots, where one charged
particle is located inside the quantum disk, whereas the other one is localized outside, enables to
reveal A-B oscillations with increasing magnetic field ~\cite{gov97,jan01}. According to this
prediction, the optical A-B effect has only recently been observed for polarized $e-h$ pairs in
type-II self-assembled quantum dots~\cite{govorovII04}. In such a  type-II  quantum disk, but with
finite thickness, where a single particle can be found above or below the disk, the angular
momentum transitions  in a perpendicular magnetic field were studied in~\cite{peet02}. In order to
explain why the A-B photoluminescence (PL) of neutral exciton has not yet been detected in type-I
semiconductor quantum rings~\cite{haft02}, the model \cite{gov02} needs further refinement. We
consider a two-dimensional quantum ring that has a different finite depth potential well of the
same width for the electron and  hole without making an assumption about the electron-hole spatial
separation (polarization). Our interest in the finite size effects has been motivated by the
prospect that the optical and electronic properties of self-assembled structures can be designed
(by reduction of potential depth and variation of geometry parameters) to allow for the optical
observation of the A-B quantum interference at low temperatures.

It is shown from the energy spectrum of an $e-h$ pair in a finite width quantum ring that the
photoluminescence emission exhibits a periodic disappearance with the magnetic field.  We discuss
the undetectability  of the A-B oscillations for neutral exciton as due to the very small magnetic
field intervals where the PL signal is absent. In our study, we consider strain free QRs. Such
objects, namely unstrained  GaAs quantum dots, were recently fabricated by novel self-assembly
technique~\cite{rast04,peet06}. The intervals in magnetic field, for which the photoluminescence is
suppressed, are calculated here numerically, and some related expressions have been derived
semiclasically. We do not consider Coulomb interaction between electron and hole in a finite width
QRs. In the literature the periodic disappearance of the optical signal from QRs has been mainly
studied in the extreme limits of weak and strong electron-hole
correlation~\cite{gov02,ull02,ph02,palm05,ulloa05,bart06,thomas01}. As it was shown
in~\cite{ulloa05}, the A-B effect remains only in the case of weak interaction and radial
polarization, though this is still under discussion~\cite{bart06}. The periodic change of the
ground state angular momentum in magnetic field is more pronounced in the case of interaction
screening and if the electron is assumed to be closer to the center of the QR~\cite{ulloa05}. Our
two-dimensional model  do not need in such assumption and also allows to take into account the
penetration of magnetic field into the material of ring, which is common for all experiments. In
the present work we model the quantum ring by an annulus with zero thickness along the growth
direction; this in-plane confinement is a rather characteristic feature for the semiconductor rings
produced by modern technology.

The paper is organized as follows. In Sec. I, we present the basic equations and the general
procedure to  obtain  the energy spectrum of a charged particle in a QR under a magnetic field. In
Sec. II, we calculate numerically the eigenenergies and wavefunctions by solving the secular
equation, estimate the $e-h$ pair overlap integral by employing the single-particle wave functions
and represent our results in form of phase diagrams of angular momentum transitions for rings of
various widths. In Sec. III we discuss semi-classically the tunneling through finite height hard
walls of QRs. In Sec. IV, we present our conclusions. Appendix A contains QR finite width
corrections to the Landau energy at high magnetic fields.

\section{Theoretical Model}

We consider an  electron (or hole) in a QR with a magnetic field $B$ applied  normally to the plane
of the ring. We assume the motion to be strongly limited along the $z$ direction and, therefore,
only consider the two-dimensional in-plane motion. Moreover, the Coulomb interaction between the
electron and hole is ignored in a first approximation.  The problem is thus reduced to one for two
independent particles, electron and hole,  with opposite charge and different effective masses,
confined by barriers of different height. Then the boundary problem for the $e-h$ pair  in a
magnetic field becomes exactly solvable. Spin coupling effects are also neglected.

The in-plane motion of a charged particle in a quantum ring under a perpendicular magnetic field is
limited by the boundary conditions imposed by the nanostructure, whereas the magnetic field
introduces a structure independent quantization.  Although the parabolic-like confinement potential
is one of the most commonly used approximation for some QRs ~\cite{chakra92,fuhrer01,tan95} (in
particular, because it allows the Fock-Darwin solution for  energy eigenvalues in an explicit form
~\cite{fock28}),  a realistic potential acting on the carriers might differ significantly. The
parabolic confining potential model has been used to study the magnetic-field-induced suppression
of the photoluminescence~\cite{gov02}. We assume here that a finite  hard wall potential limits the
motion of the electron and hole in the same radial range, but with different barrier heights, and
show that it causes the radial polarization of the $e-h$ pair. Another peculiarity of our approach
is the existence of unbound states: the finite size potential of the QR in our approximation allows
to take  them  into consideration as well, in contrast to the parabolic and infinite hard wall
models.

Within the above model, due to the axial symmetry, the Hamiltonian of a charged particle is
invariant under spatial rotation about the $z$ axis that passes through the center of the QR.
Therefore, it follows the conservation of the projection of the angular momentum on the $z$ axis,
$l_{z}$, for both electron and hole, and  the wave function can be generally factorized as
\begin{equation}
\label{eqn:psi} \Psi(\rho,\varphi)=\frac{e^{i
l\varphi}}{\sqrt{2\pi}} R(\rho),
\end{equation}
where $l=0,\pm 1, \pm 2 \dots$ is the quantum number determining the
angular momentum $l_{z}=l \hbar$.

The radial function $R(\rho)$ obey the Schr\"{o}dinger-like equation:
\begin{eqnarray}\label{eqn:Ham}
&& \displaystyle\left[-\frac{\hbar^2}{2
m}\frac{1}{\rho}\frac{\partial}{\partial \rho }\left(\frac{1}{\rho }
\frac{\partial}{\partial \rho }\right) +  \frac{\hbar^2}{2
m}\left(\frac{l}{\rho}\right)^{2} + \frac{\sigma
l}{2}\hbar\omega_{c} + \right.  \nonumber  \\ && \hspace{5cm} \left.
+\displaystyle\frac{m}{8} (\omega_{c} \rho)^2+V (\rho)\right]R
(\rho) = E R(\rho)\; , \qquad
\end{eqnarray}
Here $\omega_{c} =|e|~B/(m c)$ is the cyclotron frequency and, as
anticipated above, the ring-like confinement is assumed to be
described by the radial potential,
\begin{equation}
\label{eqn:V0} V(\rho)= \left\{
\begin{array}{l}
0, \quad \rho \leq \rho_{1}, \\[5pt]
-V_{0},\quad \rho_{1} \leq \rho  \leq \rho_{2} ,\\[5pt]
0, \quad \rho \geq \rho_{2},
\end{array}
\right.
\end{equation}
where $\rho_{1}~(\rho_{2})$ is the inner (outer) radius of the
quantum ring and $V_{0}\geq 0$. Here and below, to specify the
electron ($"e"$) and hole ($"h"$) parameters, we replace:
$m\rightarrow m_{e, h}$, $\sigma\rightarrow \sigma^{e, h}$ (with
$\sigma^{e}=- 1$, $\sigma^{h}=+1$), $R \rightarrow R^{e, h}$,
$E\rightarrow E^{e, h}$, $l\rightarrow l^{e, h}$,
$V(\rho)\rightarrow V^{e, h}(\rho)$, $V_{0} \rightarrow V^{e,
h}_{0}$, $\omega_{c}\rightarrow \omega^{e, h}_{c}$.

As it is known \cite{landau}, the radial  solution  to equation~(\ref{eqn:Ham}) is expressed in
terms of the confluent hypergeometric functions $M(a, b, z)$ and $U(a, b, z)$ \cite{abramowitz}.
Their (third) argument, the dimensionless variable $z= \frac{\rho^2}{2 l_{m}^2}$, is the magnetic
flux through the disk of radius $\rho$ in units of the elementary magnetic flux. The magnetic
length scale $l_{m}=\sqrt{ \frac{\hbar c}{|e|~B}}$ is the natural length to measure the magnetic
field dependence of the QR with different radii.

The energy spectrum of the particle can be written in a form similar to the Landau levels
expression \cite{landau}:
\begin{equation}
\label{eqn:nL} E_{\nu,l}= \hbar\omega_{c}\left(\nu+\frac{\sigma l +|l|+1}{2}\right)=
\hbar\omega_{c} \left(n_{L}+\frac{1}{2}\right),
\end{equation}
where $n_{L}$ is the principal quantum number. We count the energy from  the bottom of the
potential well. The lowest subband of energy
$\hbar\omega_{c}^{e,h}\left(\nu^{e,h}+\frac{1}{2}\right)$ corresponds  to strictly non-positive
(non-negative) values of the magnetic quantum numbers $l^{e}$ ($l^{h}$). Although for a free charge
$n_{L}$ is an integer, for a restricted motion in the plane the radial quantum numbers $\nu$ (and
therefore, $n_{L}$) become non-integer, and depend on the angular momentum quantum numbers $l$, by
virtue of the boundary conditions. In order to determine the eigenenergies one must impose that the
radial eigenfunction $R(\rho)$ satisfies the continuity conditions at the interior and exterior QR
boundaries, i.e. in the limits $\rho\to\rho_{1}$ and $\rho\to\rho_{2}$. The boundary conditions
lead to a homogeneous system of linear equations for the integration constants. By making use of
the determinant properties, we arrive at the following form of the secular equation for the allowed
single particle energy eigenvalues:
\begin{equation}
\label{eqn:arr} \left|
\begin{array}{cccc}
M(-\nu_{ex},b;r_{1}^{2}/2) & U(-\nu,b;r_{1}^{2}/2) & M(-\nu, b;r_{1}^{2}/2)~&~{0} \\[5pt]
M^{\prime}(-\nu_{ex},b;r_{1}^{2}/2)~&U^{\prime}(-\nu,b;r_{1}^{2}/2)~& M^{\prime}(-\nu,b;r_{1}^{2}/2) \, & 0 \\[5pt]
0 & U(-\nu,b;r_{2}^{2}/2) ~&~M(-\nu,b;r_{2}^{2}/2)~&~ U(-\nu_{ex},b;r_{2}^{2}/2) \\[5pt]
0 & U^{\prime}(-\nu,b;r_{2}^{2}/2)~&
~M^{\prime}(-\nu,b;r_{2}^{2}/2)&~U^{\prime}(-\nu_{ex},b;r_{2}^{2}/2)
\end{array}
\right|=0 .
\end{equation}
Here $b=1+|l|$, $r_{1,2}=\rho_{1,2}/l_{m}$, the primes denote derivatives on the radial coordinate,
and
\begin{equation}
\label{eqn:delta}  \nu_{ex}=\nu-\delta^{-1}=\nu-\frac {V_{0}}{\hbar\omega_{c}}.
\end{equation}
The  quantum numbers  $\nu$ and $\nu_{ex}$ characterize the motion interior and exterior of a ring
of width $W=\rho_{2}-\rho_{1}$. The roots $\nu$ of the secular equation (\ref{eqn:arr}) are
calculated numerically for each value of $|l|$, and the corresponding energy eigenvalues are
obtained from (\ref{eqn:nL}).

An important observation is that the secular equation depends on the
particle parameters only through the ratio $\hbar\omega_{c }/V_{0}$.
Therefore, the solutions of the secular equation and, thus, the wave
functions  are exactly the same for electrons and holes if the
following relationship between the potential and masses is
fulfilled:
\begin{equation}
\label{eqn:scale} V_{0}^{h}~m_{h}=V_{0}^{e}~m_{e}.
\end{equation}

The importance of the above equality lies in the fact that one may consider equation
(\ref{eqn:scale}) as the condition for the absence of $e-h$ radial polarization. Otherwise, when
equation (\ref{eqn:scale}) is not obeyed, the eigenenergies and eigenfunctions for the electron and
hole are different and radial polarization occurs.

\section{Numerical results for energy spectra in QRs}
\begin{figure}
\begin{center}
\includegraphics[scale=1.00]{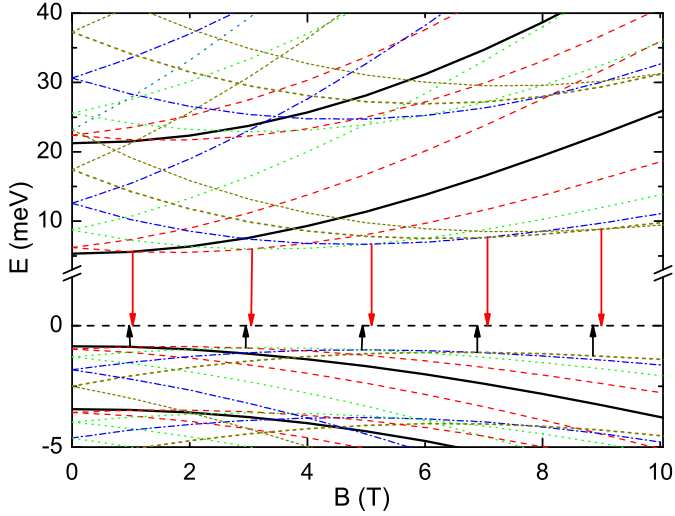}
\caption{\label{fig:e-hen} Electron and hole energy  levels versus applied magnetic field in a
potential well of $\rho_1=15$~nm, $\rho_2=40$~nm, $V_{0}^{e}=50$~meV, and
$V_{0}^{h}=V_{0}^{e}/3=17$ meV.  The intervals for existence of dark $e-h$ pair are delimited by
the nearest arrows.}
\end{center}
\end{figure}

The energy spectra have been studied systematically as a function of magnetic field in nanorings of
varying geometry. We have  calculated the energy spectra for both electron and hole in  QRs with
various confinement potentials in the range  $V_{0}^{e}=50-500$~meV, $V_{0}^{h}=17-170$ meV,
respectively. We show only the results for the case $V_{0}^{e}=50$ meV, $V_{0}^{h}=17$ meV because
the dependence of period of the ground state A-B oscillations on magnetic field has been found very
much the same for both small and large confinement energies. For definiteness, the quantum ring is
assumed to be made of GaAs material embedded in AlGaAs, and the effective masses are accordingly
taken to be $m_e=0.0667~m_{0}$, $m_h=0.5~m_{0}$, $m_{0}$ being the free electron mass.

In Figure \ref{fig:e-hen}  we show the dependence of the electron and hole energy as a function of
magnetic field for several quantum numbers $l$.  We plot the calculated energy levels versus
magnetic field for positive and negative $l$ states (whose energies converge to each other as the
magnetic field goes to zero).
\begin{figure}
\begin{center}
\includegraphics[scale=1.00]{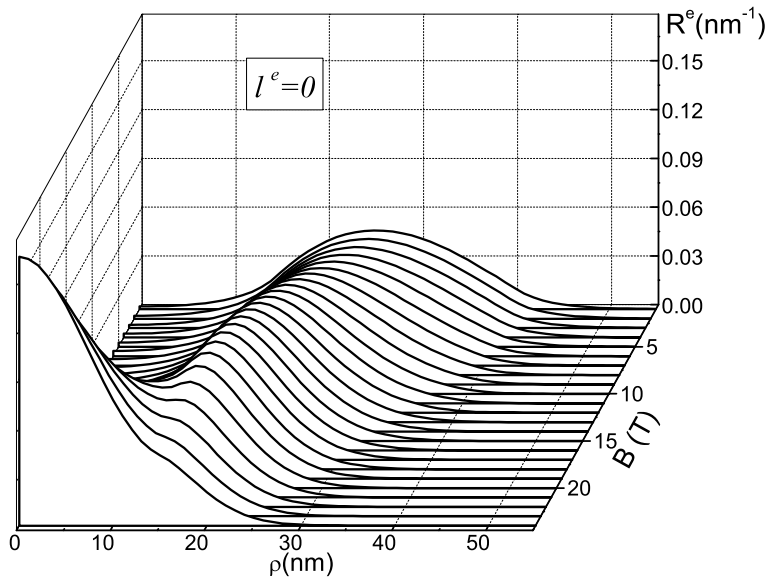}
\caption{\label{fig:wf} The evolution of the radial eigenfunction for the singlet electron,
$l^e=0$, with increasing magnetic field in a ring with the same parameters as in Fig.
\ref{fig:e-hen}.}
\end{center}
\end{figure}
From the energy level dependence with the magnetic field (see Fig. \ref{fig:e-hen}), we can observe
how the angular momentum quantum number corresponding to the lowest energy state varies with
increasing field. The intersections of the lowest energy levels indicate transitions with $|\Delta
l|=1$ at some  specific values of the magnetic field, marked by arrows in Fig. \ref{fig:e-hen};
hence the ground state energy curve consists of arch pieces crossing at these values.

At this point, we would like to note that, since for the electron in GaAs quantum dots the value of
the g factor is equal to $0.17$ \cite{okuno}, the spin splitting energy  value ($g\mu_{B} B$, where
$\mu_{B}$ is the Bohr's magneton)  is  of $0.24$  meV  at $B=25$ T. Thus it looks reasonable to
neglect, in the first approximation, the difference in the energy between the spin up and down
states.  The Zeeman gap energy was also found to be negligibly small for magnetoexcitons in
unstrained GaAs/AlxGa(1-x)As \cite{peet06}.

The above nearly periodic dependence of the energy on magnetic flux, $\Phi=\pi~\rho^2B$, through
the average area of a thin annulus is a manifestation of the A-B effect~\cite{AB59}. Indeed, the
energy of a charged particle with mass $m$ in a 1D ring of radius $\rho$ is
\begin{eqnarray}
\label{eqn:ab} E = \frac{\hbar^{2}}{2 m \rho^{2}}
\left(|l|-\frac{\Phi}{\Phi_{0}}\right)^{2},
\end{eqnarray}
where $\Phi_{0}=2\pi \hbar c/ |e|$  is the elementary magnetic flux quantum.  Furthermore, the
effect of the penetration of magnetic field into the ring itself reveals in a non parabolicity of
the energy levels, as shown in Fig.~\ref{fig:e-hen}, and in the condensation of the ground state
energy to the Landau level at high fields \cite{robnik86}.

Once the eigenvalue problem has been solved, the explicit eigenfunctions obeying the boundary
conditions given  can also be calculated. Figure \ref{fig:wf} shows the radial eigenfunction   for
the singlet electron state, $l^{e} = 0$, as magnetic field increases. Figure \ref{fig:wf} also
illustrates that, when magnetic field increases from zero, the maximum of $R^e$ evolves from inside
of the QR to the center of the ring. In the low field region, the wave function is nearly symmetric
(peaked at the average radius of the ring), while for high fields, the maximum of $R^e$ is peaked
at $\rho=0$ and warped near the inner boundary in the high field region. With increasing magnetic
field the probability of finding the singlet electron moves from the confinement region to the QR
center through the inner barrier, whereas the ground state (not shown) associated with some
$l^{e}\neq 0$ localizes away from the QR edges, and approaches to the energy of the Landau level.

In Figure \ref{fig:e-hen} the magnetic field values corresponding to the successive changes of $l$
(marked by arrows) depend on the effective mass and height of the confinement potential and thus
they are different for the electron and hole if equation (\ref{eqn:scale}) is not fulfilled. Figure
\ref{fig:e-hen} suggests that there exist a quite narrow but finite ranges of magnetic field values
where the total angular momentum of the $e-h$ pair, $L_{z}=l^{e}+l^{h}$, changes abruptly from
$L_{z}=0$ to a  non vanishing value. This has an impact on the oscillator strength of the $e-h$
pair. Indeed, the photoluminescence intensity is proportional to the square of the overlap integral
\begin{eqnarray}
\label{eqn:per} \int \Psi_{e}(\rho,\varphi )\Psi_{h} (\rho,\varphi )
d\vec r=\Xi~\delta_{0,L_{z}}, \qquad \Xi=\int\limits_{0}^{\infty}
R_{\nu^{e }, l^{e}}(\rho) R_{\nu^{h},l^{h}}(\rho)\rho d\rho.
\end{eqnarray}

It follows from equation (\ref{eqn:per}) that the optical emission process must obey the selection
rule $L_{z}=0$, i.e. the exciton emission is only possible for the singlet  state of the $e-h$ pair
in which $l^{e}=-l^{h}$ \cite{gov97}.  Thus, the electron-hole recombination is forbidden within
the intervals where the so-called dark exciton $(L_{z}\not=0)$ becomes the ground state. These
intervals correspond to the domains between nearest arrows in Fig. \ref{fig:e-hen}. Since the
overlap integral $\Xi$ has a very weak dependence on $B$ (as obtained from our numerical
calculations), these are the only domains where e-h recombination is forbidden. In the particular
case of a QR with infinite barriers, equation (\ref{eqn:scale}) holds trivially, and the photon can
be emitted for any value of $B$.

\begin{figure}
\begin{center}
\includegraphics[scale=0.60]{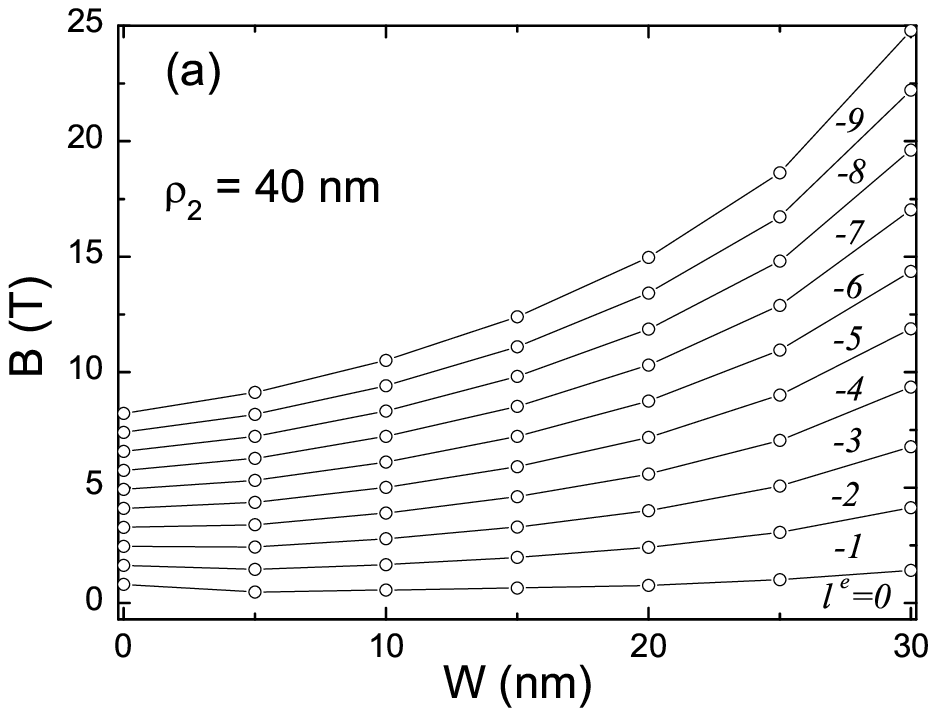}
\includegraphics[scale=0.60]{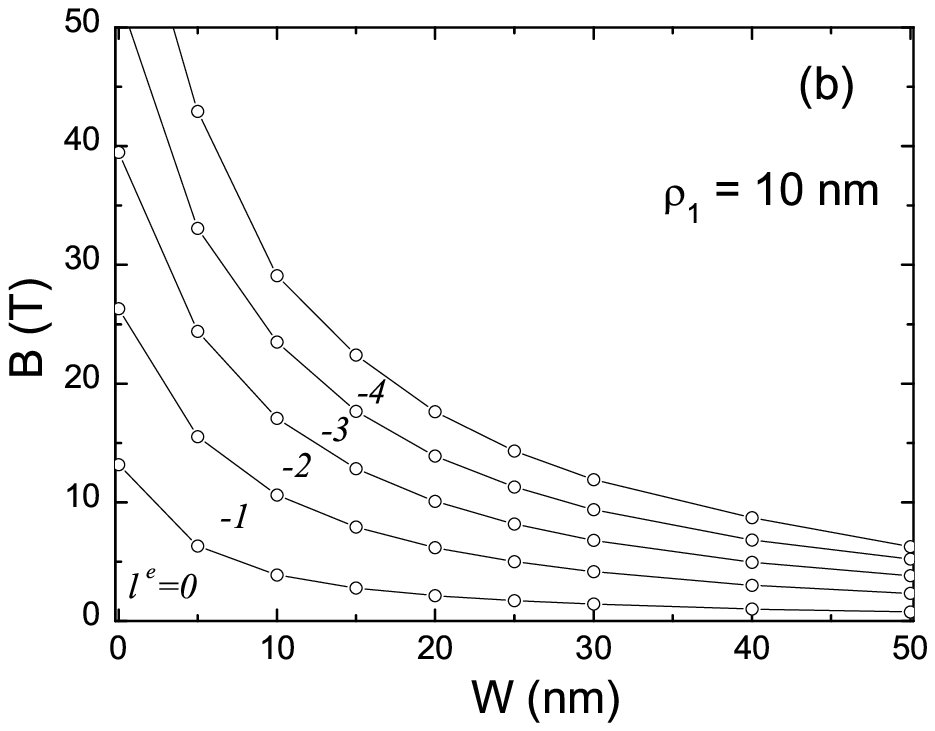}
\caption{\label{fig:r2} Phase diagrams showing the electron angular
momentum transitions, $l^{e}\rightarrow l^{e}-1$, in a QR of the
$V_{0}^{e}=50$ meV, for fixed outer radius $\rho_{2}$=40 nm (a), and
fixed inner radius $\rho_{1}$=10 nm (b).}
\end{center}
\end{figure}
Figures \ref{fig:r2} (a) and (b) summarize our numerical results concerning the angular momentum
transitions in the form of phase diagrams of magnetic field versus ring width~\cite{jan01}. In both
figures, the flux of the magnetic field must approach an integer number of the elementary flux for
the 1D case ($W\to 0$), in accordance to the well known expression (\ref{eqn:ab}).

For fixed $\rho_2$, the magnetic field period of angular momentum transitions increases with the
increasing width of the QR, as illustrated in Fig.~\ref{fig:r2}(a). As $\rho_{1}$ vanishes, i.e. $W
\rightarrow \rho_{2}$ the period of A-B oscillations increases asymptotically to infinity. The
change in topology of  the confining potential results in the disappearance of the angular momentum
transitions in the case of a quantum disk. On the other hand, as Fig.~\ref{fig:r2}(b) shows, for a
fixed $\rho_1$, as the  width increases the magnetic field period decreases monotonically. In the
large width limit, the A-B  flux tends asymptotically to the characteristic periods of a quantum
antidot system, see for example~\cite{peet02}.

Surprisingly, our calculations (not shown here) indicate that the
period of A-B oscillations in a magnetic field is almost not
affected by a tenfold increase of the confining potential. This was
verified for various radii $\rho_1, \rho_2$, but keeping the fixed
ratio $V_0^e/V_0^h=3$, for both electron and hole. In either
potential the phase diagrams are nearly indiscernible due to the
fact that the scaling equation (\ref{eqn:scale}) is approximately
satisfied. The magnitude of the potential influences, for instance,
the range of magnetic field for which the bound ground state
transforms into an unbound and viceversa (see below).

Notice that the ultimate reason for the disappearance of the photoluminescence signal from a ring
is the selection rules following from the conservation of the total  angular momentum  of the {\it
e-h} pair. Therefore, it is tempting to speculate that our results on the periodicity of the
disappearance of the optical signal with the magnetic field, as based on the rotational symmetry of
the 2D system, remain valid,  at least qualitatively, in the presence of an {\it e-h} Coulomb
interaction. As it was shown in~\cite{muller93,chakra94}, for a system of interacting electrons the
magnetization is nearly the same as for the non-interacting electrons  due to conservation of
angular momentum in the 2D quantum rings. The numerical results of~\cite{chakra94} demonstrate
that, in the lowest Landau level, the electron-electron Coulomb interaction shifts the
non-interacting energy spectrum to higher energies. Although it is commonly believed that the A-B
effect exists also for excitons, in the weakly interacting regime the very recent studies show some
discrepancy on this issue~\cite{ulloa05,bart06}.  In this regime strong A-B oscillations with  the
magnetic field remain~\cite{ulloa05} (also our approach supports this statement). In the weakly
interacting regime, in the first order of perturbation theory, the electron-hole Coulomb
interaction shifts rigidly by a negative constant the non-interacting energy spectrum to lower
energies~\cite{bart06}. The natural continuation of the present research is the investigation of
the effect of the {\it e-h} interaction on the magneto-optical transitions by complete
diagonalization of the Hamiltonian, see {\it e.g.}~\cite{chakra94}, using our electron hole
non-interacting basis. This might appear to be useful as complementary to the other
approaches~\cite{palm05,ulloa05} being under intensive investigations now.  In the strongly
correlated regime, the only remaining A-B effect manifests itself in oscillating excited
states~\cite{ulloa05,bart06}. A more complete study of effects of the Coulomb interaction for
polarized {\it e-h} pair in continuous 2D model of QRs beyond perturbation theory is still to be
done.

As examples of possible qualitative effects coming from the turning on an interaction, we refer
also on  exact non-perturbative results obtained for discrete-lattice rings, which in contrary to a
Galilean-invariant system, see {\it e.g.}~\cite{muller93,romer95}, yield such dependence of the
magnetization on the correlation. It was shown that, as the interaction strength varies, the ground
state gaps open and close, see for example~\cite{romer95}, while only for gapless excitations, the
amplitude of the A-B oscillation is not exponentially small~\cite{zvyag93}. In the Hubbard chain,
the strong correlation results in the changes of magnitude and period of the ground state energy
oscillations, see {\it e.g.}~\cite{zv&kr95}; the period of the A-B oscillations, depending on
magnetization of 1D ring of the fermion system with attraction, alters from integer to half integer
of $\Phi_{0}$~\cite{zvyag93}. The extra frequencies to the fundamental Aharonov-Bohm frequency of
the exciton oscillations were shown to appear in a ring described by the attractive Hubbard model
on the coupled concentric 2D annular lattices~\cite{palm05}.  In our paper we assume that the QR is
an ideal 2D ring and that the roughness of boundary shape nor defects do not smash periodicity of
equilibrium thermodynamic quantities~\cite{byers}. It has been shown that elastic scattering does
not attenuate the A-B effect in a ring with disorder~\cite{buttiker}.

\section{Curvature effect on tunneling through ring under a magnetic field}

In this Section we will supplement our quantum mechanical
calculations by a semi-classical determination of the magnetic field
values $B_{1}(B_{2})$ for which the electron or hole states cross
the inner (outer) ring boundaries in connection to the feasible
domain for the existence of dark excitons.

The radial polarization of carriers with different masses in a
finite confining potential is essentially a quantum mechanical
effect. It appears due to the fact that the probability of tunneling
through a potential barrier for one charged particle exceeds the
probability for the other.

The semi-classical WKB theory says that the quantum energies for 2D
motion of a charged particle in a cylindrically symmetrical
potential and under a magnetic field are defined by the roots of the
equation~\cite{landau}
\begin{eqnarray}
\label{eqn:eff} \pi~(\nu+\gamma)=\sum_{i}
\int\limits_{\hat{r}_{i}}^{\hat{r}_{i+1}}
\sqrt{\frac{E}{\hbar\omega_{c}}-U_{eff}(l,r,\delta)}~dr \, ,
\end{eqnarray}
where $\gamma$ is a constant and the integral is over the
classically allowed region (where the square root has a positive
argument) limited by a couple of turning points (${\hat{r}_{i}},
{\hat{r}_{i+1}}$). To define this region we consider the minima of
the effective potential $U_{eff}(l, r,\delta)$ at a given magnetic
field.

In our case, the effective  potential is defined by the finite confinement potential, which takes
zero values outside the QR, equation (\ref{eqn:V0}), combined with the quantum-mechanical
centrifugal potential barrier $l^2/r^2$,  and with the magnetic potential $r^2/4-\sigma l$:
\begin{equation}
\label{eqnarray:W} U_{eff}(l, r,\delta)= \left\{
\begin{array}{cc}
(-\sigma l/r+ r/2)^2, & r ~\leq r_{1},\\[5pt]
- 2/ \delta +(-\sigma l/ r+ r/2)^2, & { r_{1}~\leq~ r ~\leq~ r_{2}} \\[5pt]
(-\sigma l/ r+ r/2)^2, & { r ~\geq~ r_{2}}\; ,
\end{array}
\right.
\end{equation}
where $r=\rho/l_m$, $r_{1}$ can be considered as the inner radius of the ring or antidot, $r_{2}$
as the outer radius of ring or disk in units of magnetic length, and  $1/\delta$ is the number of
Landau levels inside the potential, equation (\ref{eqn:delta}).

We shall substitute $\nu\rightarrow \nu^{e, h}$, $\delta\rightarrow
\delta^{e, h}$, $m \rightarrow m_{e, h}$, $E\rightarrow E^{e, h}$,
$V_{0}\rightarrow V_{0}^{e, h}$, $l \rightarrow l^{e, h}$,
$\sigma=\mp 1$, when discussing the electron and the hole,
respectively. The effective potential depends on the absolute value
of the magnetic quantum number $\sigma l=|l|$ for the electron as
well as for the hole in the ground state. The integration allows us
to obtain the transcendental equation for the energy as explained in
Appendix A.

At high magnetic field, the appearance of the bulk states with
Landau energy in the QR (Landau condensation)~\cite{robnik86}
results in a factorization of the secular equation into the
characteristic equation for the antidot of radius $\rho_{1}$ and
that for the circular disk of radius $\rho_{2}$ \cite{halp81}. That
allows us to consider separately the states of a charged particle
near the outer and inner boundaries of ring confining potential.

\begin{figure}
\begin{center}
\includegraphics[scale=1.00]{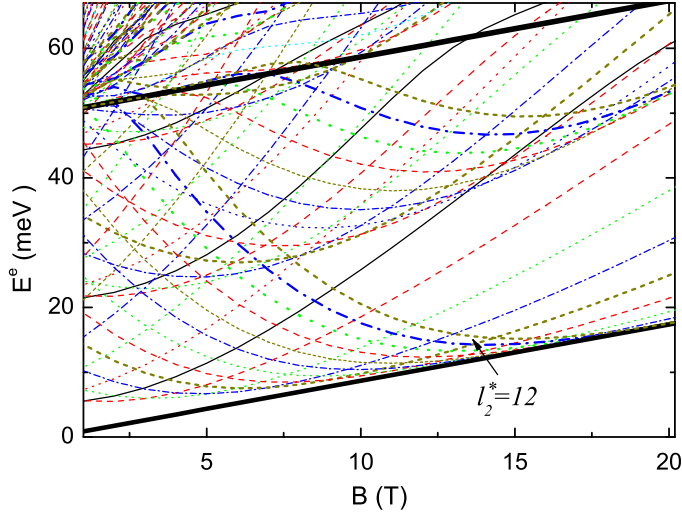}
\caption{\label{fig:l2} The calculated electron energy levels as a
function of magnetic field for a QR of $\rho_1$=15 nm, $\rho_{2}$=40
nm, and $V_{0}^{e}=50$ meV. The solid line corresponds to $l=0$,
dashed lines to $|l|=1, 5, 9$, dotted lines to $|l|=2, 6, 10$,
dash-dotted lines to $|l|=3,7,11$, short-dotted lines to
$|l|=4,8,12$. The two straight parallel thick lines depict the
lowest Landau levels inside ($E_{0,0}^e$) and outside
($E_{0,0}^e+V_{0}^e$) the QR confinement.}
\end{center}
\end{figure}

Firstly, we shall consider negative (positive) $l$ states of the
electron (hole) in a circular disk confining potential. The
effective potential  derivative vanishes at $r_{l}=\sqrt{2|l|}$,
where the probability of finding the charge in the eigenstate
$\Psi_{\nu,l}$ has its maximum. If $r_{l}<r_{2}$, there is one well
effective potential with the minimum value $U_{eff}(l,
r_{l},\delta)=- 2/\delta$ at $r=r_{l}$. If $|l|$ is sufficiently
large, the probability of finding a charged particle out of the
confinement region, $r_{l} >r_{2}$ appears to be finite. Therefore,
there emerges a double well effective potential for the states
because of the appearance of two minima: the inside minimum is just
at the outer boundary, $r-r_{2}\rightarrow 0^-$ (because of the
step-like change of the confinement potential) and the outside
minimum is at $r_{l}>r_{2}$. We define the magnetic quantum number
$l_{2}^{*}$ as the number such that for $|l| \leq l_{2}^{*}$ the
minimum of $U_{eff}$ in the QR, at $r-r_{2}\rightarrow 0^-$, is
deeper than the minimum  out it, at $r=r_{l}$. These minima become
nearly equal when the angular momentum quantum number reaches the
value $l_{2}^{*}$ (up to its modulus)
\begin{eqnarray}
\label{eqnarray:m2} U_{eff}(l_{2}^{*},r_{2},\delta) \approx
U_{eff}(l_{2}^{*},~r_{l},\delta),\qquad
r_{l} > r_{2} \nonumber \\[10pt]
l_{2}^{*}=\left[\frac{\Phi_{2}}{\Phi_{0}}+\sqrt{2
\frac{\rho_{2}^{2}}{\hbar^{2}} V_{0} m} \,\right].
\end{eqnarray}
In equation (\ref{eqnarray:m2}) $\Phi_{2}$ is the magnetic flux threading the disk of  radius
$\rho_{2}$, and the square brackets denote the integer part of the  wrapped expression. From
equation (\ref{eqnarray:m2}) we estimate  $l_{2}^{*}~(B=1~T)$ to be about $12$ for the electron in
the potential $V_{0}=50$ meV. Indeed, the curve for the energy of the electron state with $l^e =10$
starts to split  from the Landau level outside the ring  ($E_{0,0}^e+V_{0}^e$) in a low magnetic
field, as Fig.~\ref{fig:l2} shows. Therefore, the energy level for $|l|\geq l_2^{*}$, manifests an
inflection, and then approaches the Landau level inside the QR ($E_{0,0}^e$) at high magnetic
fields.
An increasing magnetic field pushes negative eigenstates for the electron to the origin, as it is
shown in Fig.~\ref{fig:wf}. As it follows from equation~(\ref{eqnarray:m2}), the maximum of
probability density of the state associated with a given $l$ localizes inside of the external QR
boundary when $B\geq B_{2}$:
\begin{eqnarray}
\label{eqn:B2} B_{2}(l) = -\frac{c\sqrt{8~m~V_{o}}}{|e|\rho_2}
+|l|\frac{\Phi_{0}}{\pi \rho_{2}^{2}}.
\end{eqnarray}
The $S$-state ($l=0$) has the global minimum of the effective
potential fixed at the center of the disk at arbitrary magnetic
field. In the case of disk geometry, only this state contributes
appreciably to the ground state for low fields, whereas the energies
of the nonzero angular momentum states just merge successively into
the Landau level  of the $l=0$ state in high magnetic field
\cite{robnik86}. In the case of annular geometry, with increasing
magnetic field the states consecutively tunnel via the inner barrier
and, due to its contribution, the magnetic quantum number $|l|$ of
the ground state increases starting from zero.

\begin{figure}
\begin{center}
\includegraphics[scale=0.60]{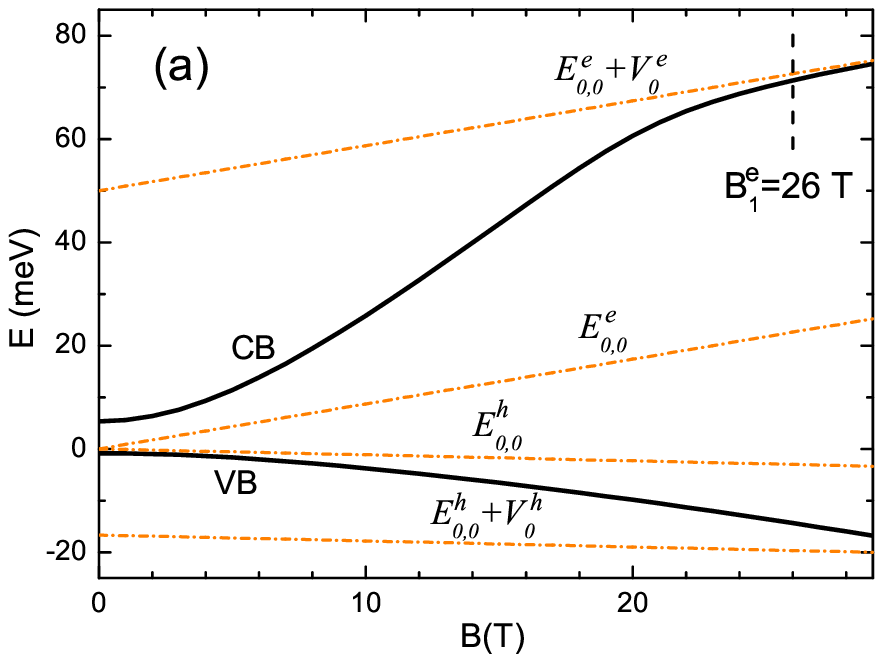}
\includegraphics[scale=0.60]{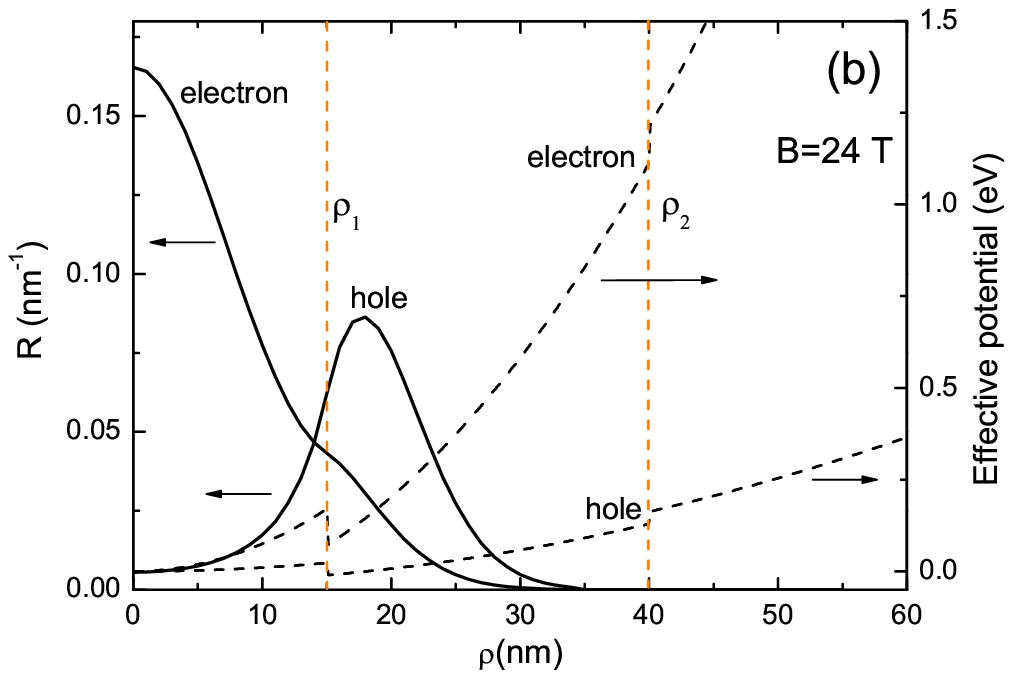}
\caption{\label{fig:entrance} (a) The calculated ground state
electron and hole energies versus  magnetic field for the QR of
$\rho_1$=15 nm, $\rho_{2}$=40 nm, and $V_{0}^{e}=50$ meV,
$V_{0}^{h}=17$ meV, respectively. (b) Corresponding electron and
hole radial wave functions at a magnetic field of B=24 T, in the
same QR. The figure also shows the electron and hole potentials
(right scale).}
\end{center}
\end{figure}

Secondly, we address  the  problem of inward tunneling  of the $l$
states in the antidot confining potential. There is a double well
for the $S$-state and the probability density has two maxima binding
to the minima of the effective potential. In a weak magnetic field
the local minimum at $\rho=0$ is higher than the step-wise minimum
at the inner boundary of the ring, $\rho- \rho_{1}\to 0^+$. Under a
certain magnetic field, $B_{1}(l=0)$, when the confinement barrier
becomes equal to the characteristic magnetic energy, $V_{0}
=\frac{1}{8} m \omega_{c}^2 \rho_{1}^2$, the $S$-state penetrates
into the ring opening, as Fig.~\ref{fig:wf} shows. In the case of
arbitrary $l$, the two wells of effective potential level at the
threshold value of magnetic field
\begin{equation}
\label{eqn:B1} B_{1}(l) = \frac{c\sqrt{8 m V_0}}{|e|\rho_1}+
|l|\frac{\Phi_{0}}{\pi \rho_{1}^{2}},
\end{equation}
and the $l$ state appears in the ring opening if $B \geq B_{1}(l)$. From equation (\ref{eqn:B1}) we
estimate  $B_{1}^e(l^e=0)=25.94~T$ that correlates well with the $E(B)$ dependence shown in
Fig.~\ref{fig:entrance}(a). So far, our semi-classical estimates of $B_1$ and $B_2$ (or
$l_{2}^{*})$ are in good agreement with the results of quantum mechanical calculations. Taking
together the conditions for passage through the inner and outer barriers, we conclude that the $l$
state is trapped inside the ring if $B_{2}(l) \leq B\leq B_{1}(l)$. We present  in the Appendix A
the transcendental equations for the energy of the $l$ state within this range of the magnetic
field values.

The energy is a function of $\rho_{l}=l_{m}\sqrt{2|l|}$, which is the guiding cyclotron orbit
center for unrestricted motion in a plane \cite{cohen}. States with $\rho_{l}$ in the vicinity of
the boundaries of a confining potential are known as {\it edge states}. The distance from  the
orbit center of an edge state to the boundary is less than the magnetic length. There are {\it
inner} and {\it outer} edge states  oppositely circulating from the inside of the QR of the
infinite potential under a weak magnetic field,  which induce diamagnetic and paramagnetic moments,
respectively~\cite{lent91}. The inflection of the energy curves shown in Fig.~\ref{fig:l2} and
Fig.~\ref{fig:entrance}(a) can be related to the edge states of the electron and interpreted in
terms of the magnetic susceptibility. Indeed, the magnetization and the magnetic susceptibility of
the $R_{\nu, l}$ state of a single-charged particle  are defined by
 \begin{eqnarray}
\label{eqn:DL}
 M_{\nu, l}(B)=-\frac{\partial }{\partial B }E_{\nu, l}(B),\\
\chi_{\nu,l}=\frac{\partial }{\partial B }M_{\nu, l}(B).
\end{eqnarray}
The magnetic susceptibility  of the $l$ state, $\chi_{\nu,l}$ changes  sign at an inflection point
of  the curve $E_{\nu,l}(B)$. We have found, as Fig.~\ref{fig:l2}  shows, that the energy of an
eigenstate with a given $l$, beginning from $|l|=l_{2}^{*}$ reveals an inflection due to the change
in $M_{\nu, l}(B)$ on the outer boundary. The electron edge state associated with $|l| \geq
l_{2}^{*}$ circulates in counter clockwise direction (about the $z$ axis) nearby the external
boundary from outside. With increasing magnetic field the center of the wave function approaches
the ring origin and the circulation changes direction to the opposite one inside of the external
boundary  due to the scattering from it. Then, as the magnetic field increases,  this eigenstate of
positive energy beyond $\rho_{2}$, transforms into the ground state, $-|V_{0}| \leq E \leq 0$, in
an interior region of the QR. Figure~\ref{fig:l2} shows this type of inflection point on the energy
level ($M_{\nu, l}(B)>0$) for the electron. As the magnetic field increases, the energy curve is
flattened nearby the Landau level. Finally,  it tends asymptotically to the bulk level in the
interior region of the ring opening as Fig.~\ref{fig:entrance}(a) shows.  As a result, the electron
edge  state circulates  in clockwise direction along the inner boundary from inside.  The
inflection of the raising energy curve $E(B)$ shown in Fig.~\ref{fig:entrance}(a), implies the
change of the magnetic response from diamagnetic, $\chi_{\nu,l}\leq 0$, to  paramagnetic,
$\chi_{\nu,l}\geq 0$, whereas the inflection of the falling energy curve indicates the inverse
change on the outer boundary, see Fig.~\ref{fig:l2}.

The energy branch for the singlet state of the hole is also depicted
in Fig. \ref{fig:entrance}(a). The maximum of the electron wave
function is at the center of the QR. The shoulder (a second maximum
at lower fields [see Fig. 2]) at $\rho=15$ nm is due to a deep in
the electron potential, as shown in the Fig. \ref{fig:entrance}(b).
The hole wave function exposes only one peak located at the absolute
potential minimum, because $B\leq B_{1}^h(l^h=0)$, where
$B_{1}^h(l^h=0)$=41T.

For the singlet pair $L_{z}=l^e+l^h=0$  we readily derive from
Eqs.~(\ref{eqn:B2}), (\ref{eqn:B1}) the difference between  the
magnetic field values for the  the electron and the hole states when
crossing the outer and inner boundaries of the confining potential
region
\begin{equation}
\label{eqn:gap2} B_{2}^{e}-B_{2}^{h}=
+\frac{c\sqrt{8~m_{h}~V_{0}^{h}}}{|e|\rho_{2}}(1-\Gamma),
\end{equation}
\begin{equation}
\label{eqn:gap1} B_{1}^{e}-B_{1}^{h}=
-\frac{c\sqrt{8~m_{h}~V_{0}^{h}}}{|e|\rho_{1}}(1-\Gamma),
\end{equation}
where
\begin{equation}
\Gamma=\sqrt{\frac{m_{e}~V_{0}^{e}}{m_{h}~V_{0}^{h}}}.
\end{equation}
The  $+$ ($-$) sign  and the index $2$ ($1$) stand for outer (inner) boundary. We employ a value of
potential depth for the hole of approximately  one third of the electron one. We  estimate
$\Gamma\approx 0.63$ with these data, and conclude from equation~(\ref{eqn:gap2}) that, for the
$L_{z}=0$ pair at arbitrary $l$, the electron state  becomes bound  at a magnetic field $B_{2}^{e}$
which is greater that of the hole, $B_{2}^{e}\geq B_{2}^{h}$.  In contrast, the electron state
becomes unbound in the QR opening at a magnetic field value $B_{1}^{e}$ lower than the one for the
hole, $B_{1}^{(e)}\leq B_{1}^{(h)}$.  The electron passes through the barriers at the same magnetic
field as the hole if $\Gamma=1$, in agreement with the implicit condition of the absence of radial
polarization, equation~(\ref{eqn:scale}).

If we are interested in the magnetic field values where the PL disappears since total angular
momentum of magneto-exciton is not zero, it follows from the quantum mechanical calculations that
these intervals  are rather narrow for the selected parameters. Perhaps due to this, the A-B effect
has not yet been observed in optical experiments in type I semiconductor structures. Nevertheless,
from Eqs.~(\ref{eqn:gap2}) and (\ref{eqn:gap1}) one might conclude that the domain for
disappearance of the  PL signal depends on inherent and geometric parameters of self-assembled QR.
Our proposal is to extend the interval for existence of dark exciton in magnetic fields by choosing
semiconductors with parameter $\Gamma$ as smaller as possible to increase polarization and to allow
the optical observation of the neutral exciton A-B effect.

\begin{sloppypar}

\section{Conclusions}
We have explored in detail the role of the finite width effects and
potential depth on the dispersion of energy levels of a QR as a
function of magnetic field and angular momentum quantum number. The
results of the numerical calculations have been summarized in phase
diagrams for the angular momentum transitions as a function of
magnetic field and various geometry parameters. We have studied the
transitions between  the unbound states and the ground state through
the energy spectra, and proposed formulae to estimate the  magnetic
field values related to these transitions. That gives rise to a
suggestion for detection of A-B optical signal from the
electron-hole pair confined in the semiconductor quantum ring,
perpendicular to magnetic field. Taking into account the Coulomb
interaction by the complete diagonalization of the exciton
Hamiltonian is the problem for future study.

\section{Acknowledgments}
This work has been supported by the grant $No-SAB2000-0353$ from the
Ministry of Education, Science, Culture and Sport of Spain. One of
the authors, T. B., would like to thank Jos\'{e} A. de Azc\'{a}rraga for
valuable comments on the manuscript.

\end{sloppypar}
\appendix
\section{Appendix: Finite width corrections}
The Bohr-Sommerfield quantization rule determining the allowed energies of the particle in the
effective potential, equation~(\ref{eqn:eff}), on substituting $z=r^{2}/2$, takes the form
\begin{equation}
\label{eqn:zet} 2~\pi~(\nu+\gamma)=\sum_{i}
\int\limits_{\hat{n}_{i}}^{\hat{n}_{i+1}} \lambda(z)~\frac{d~z}{z},
\qquad
\end{equation}
where $z_{1}\leq \hat{n}_{i}$, $\hat{n}_{i+1}\leq z_{2}$, and
\begin{equation}\label{l=l=}
\lambda(z)\equiv \lambda(z; z_{1}, z_{2})=\sqrt{(z-z_{1})(z_{2}-z)}.
\end{equation}
Here the integration range is split into two intervals for the
two-well effective potential for the energy  between its two minima
values, whereas there is one interval for the one-well potential.

Evaluating the integral yields
\begin{equation}
\label{eqn:int}
\begin{array}{lcl}
\kappa(z; z_{1}, z_{2})& = & \displaystyle\int \lambda(z; z_{1}, z_{2})~\frac{d~z}{z}  \\
& = & \displaystyle\lambda(z)-\bar{z}
\arctan\frac{\bar{z}-z}{\lambda(z)}+ \sqrt{z_{1}~z_{2}}
\arctan\frac{z_{1}~z_{2}-\bar{z}~z}{\lambda(z)}.
\end{array}
\end{equation}
We refer to the electron throughout this Appendix, and introduce the
dimensionless  energy variable
$\epsilon=2~\frac{E}{\hbar\omega_{c}}$, so that
\begin{equation}
\label{eqn:raiz} z_{1}~z_{2}=l^{2} ,\qquad
\bar{z}=\frac{1}{2}(z_{1}+z_{2})=\epsilon-l\; . \qquad
\end{equation}

We consider further the states bound to the lowest minimum inside the confining potential within
the range, $B_{2}(l) \leq B \leq B_{1}(l)$.  The limits of integration are $\hat{n}_{1}= N_{1},~
\hat{n}_{2}=N_{2}$, where   $N_{1,2}=\Phi_{1,2} /\Phi_{0}$ is the number of magnetic fluxes through
a circular disk of radius $\rho_{1,2}$, respectively. In the limiting case of  a narrow ring,
$W=\rho_{2}-\rho_{1}<<l_{m}$,  we obtain equation~(\ref{eqn:ab}) for the energy spectrum of the
particle in the zero-th order approximation on $ W/\rho_{1}<<1$, and  the results of \cite{kost92}
in the next order. In the case of a wide ring, $W>>l_{m}$, when the left and right turning points
are $\hat{n}_{1}= z_{1},~ \hat{n}_{2}=z_{2}$, and, $\gamma=1/2$,  we recover the bulk Landau
energies equation~(\ref{eqn:nL}).

At fixed high magnetic field the bulk Landau states have the inner
and outer edge states from both sides of the QR boundary. For the
edge states, the spectra can be re-written in the form which is
characteristic of the 2D oscillator
\begin{eqnarray}
\label{eqn:perim}
E_{\nu,l}= 2~\hbar~\omega_{c}(\nu+\frac{l+|l|}{4}+\frac{1}{2})+f_{\nu,l}.
\nonumber
\end{eqnarray}

At high magnetic field, for the inner edge states characteristic of the antidot, the integration is
over the interval $[N_{1}, z_{2}](N_{1} \geq z_{1})$ and from equation~(\ref{eqn:zet}) we arrive
at:
\begin{eqnarray}
\label{}
 2~\pi~(\nu+\gamma)=-\lambda(N_{1})+\frac{\pi}{2}(\epsilon-l-|l|)+\nonumber \\+(\epsilon-l) \arctan\frac{\epsilon-l-N_{1}}{\lambda(N_{1})}
 -|l| \arctan\frac{l^{2}-(\epsilon-l)~N_{1}}{\lambda(N_{1})} ,
\end{eqnarray}
where $\lambda$ is defined by (\ref{l=l=}) with $z_{1,2}$ from
(\ref{eqn:raiz}),
\begin{eqnarray}
 \lambda(N_{1})=\sqrt{-N_{1}^{2}+2(\epsilon-l) N_{1}-l^{2}}\, .\nonumber
\end{eqnarray}
Then, for this case  $f_{\nu,l}$ is given by
\begin{eqnarray}
\label{eqn:dia}\hspace{-0.8cm}
f_{\nu,l}^{AD}=\frac{\hbar~\omega_{c}}{\pi}[\lambda(N_{1})-(\epsilon-l)
\arctan\frac{\epsilon-l-N_{1}}{\lambda(N_{1})}
 +|l| \arctan\frac{l^{2}-(\epsilon-l)~N_{1}}{\lambda(N_{1})}].
\end{eqnarray}
It turns out, from equation (\ref{eqn:dia}), that the inner perimeter correction to the
magnetization is a diamagnetic one, i.e. $-\frac{\partial }{\partial B }f_{\nu,l}^{AD}(B) \leq 0$.

For the outer edge states peculiar to the disk geometry, the
integration is over the interval $[z_{1}, N_{2}] (N_{2}\leq z_{2})$
and $f_{\nu, l}$ is
\begin{eqnarray}
\label{eqn:para}\hspace{-0.9cm}
f_{\nu,l}^{D}=\frac{\hbar~\omega_{c}}{\pi}[-\lambda(N_{2})+(\epsilon-l)
\arctan\frac{\epsilon-l-N_{2}}{\lambda(N_{2})}
-|l|\arctan\frac{l^{2}-(\epsilon-l)~N_{2}}{\lambda(N_{2})}].
\end{eqnarray}
The  outer perimeter correction to the magnetization proves to be a
paramagnetic one, $-\frac{\partial }{\partial B }f_{\nu,l}^{D}(B)
\geq 0$, while the inner perimeter correction is diamagnetic one.

The correction to the energy from the inner, $f_{\nu, l}^{AD}$, and
outer, $f_{\nu, l}^{D}$, boundaries of the ring are of opposite sign
(diamagnetic and paramagnetic shift of the Landau state), though,
overall, the surface states increase the bulk Landau energy, since
that is a universal feature of an enclosure \cite{bk:courant}.

The  intersections of the set of ascending energy levels for the inner edge states, equation
(\ref{eqn:dia}), with the set of descending energy levels for the outer edge states, equation
(\ref{eqn:para}), allow to determine the  magnetic field values for angular momentum transitions
where the energy is degenerate.

\bibliographystyle{elsart-num}

\newpage
\listoffigures
\end{document}